%% file: main.tex
\documentclass[twoside]{article}

\usepackage[accepted]{aistats2024}
\usepackage[round]{natbib}

\usepackage{subcaption}
\usepackage{hyperref} 
\usepackage{booktabs}
\usepackage{adjustbox}
\usepackage{multirow}
\usepackage{graphicx}
\usepackage{pifont}
\usepackage{amsmath}
\usepackage{amssymb}
\input{figures/bind_ops.tex}
\newcommand{\xmark}{\ding{55}}
\usepackage{tikz}

\begin{document}

\twocolumn[

\aistatstitle{Holographic Global Convolutional Networks for Long-Range Prediction Tasks in Malware Detection}

\aistatsauthor{ Mohammad Mahmudul Alam\textsuperscript{1}  \And Edward Raff\textsuperscript{1,2} \And  Stella Biderman\textsuperscript{2} \AND Tim Oates\textsuperscript{1} \And James Holt\textsuperscript{3} }

\aistatsaddress{\textsuperscript{1} University of Maryland, Baltimore County \And \textsuperscript{2} Booz Allen Hamilton \And \textsuperscript{3} Laboratory for Physical Sciences } ]

\runningauthor{Alam, Raff, Biderman, Oates, Holt}

\begin{abstract}
Malware detection is an interesting and valuable domain to work in because it has significant real-world impact and unique machine-learning challenges. We investigate existing long-range techniques and benchmarks and find that they're not very suitable in this problem area. In this paper, we introduce Holographic Global Convolutional Networks (HGConv) that utilize the properties of Holographic Reduced Representations (HRR) to encode and decode features from sequence elements. Unlike other global convolutional methods, our method does not require any intricate kernel computation or crafted kernel design. HGConv kernels are defined as simple parameters learned through backpropagation. The proposed method has achieved new SOTA results on Microsoft Malware Classification Challenge, Drebin, and EMBER malware benchmarks. With log-linear complexity in sequence length, the empirical results demonstrate substantially faster run-time by HGConv compared to other methods achieving far more efficient scaling even with sequence length $\geq 100,000$.
\end{abstract}

\section{Introduction}
Ever since the transformer \citep{transformer} revolutionized natural language processing research \citep{brown2020language,devlin2018bert,raffel2020exploring}, significant attention has been paid to the quadratic cost of increasing sequence length. While traditional academic benchmarks tend to not require sequence lengths beyond 4096, many real-world applications such as multi-round chat \citep{MosaicML2023Introducing,yao2023deepspeed}, biological sequence modeling \citep{ahdritz2022openfold,avsec2021effective,dalla2023nucleotide,jumper2020alphafold,lin2022language}, and analyzing computer programs \citep{hrrformer,bloomz,roziere2023code} do. The unique challenges, data, and sequence dynamics that occur within each application can have a significant effect on what techniques work well, which is not well elucidated within the current Transformer literature. 

In this paper we are concerned with malware classification using byte-level representations of executables \citep{raff2017malware}, a task that can require sequence lengths of up to 200 million in common real-world scenarios. Though we are not able to process this extreme length in its entirety, we focus on it as an important research direction to test and develop algorithms for long-sequence task modeling. In particular, we find that some popular benchmarks from natural language processing are not well correlated with improvement in malware detection tasks. Thus, we find it necessary to develop new architectures, which we do by incorporating aspects of classical neuro-symbolic methods like the Holographic Reduced Representation (HRR)~\citep{hrr}.

\subsection{Malware Detection}
Two predominant types of malware detection tasks exist: distinguishing malicious programs from benign and distinguishing a known malicious file into unique families of malware. Both of these tasks are relevant to real-world cyber security and are complicated by the long-range interactions, spatial and non-spatial locality, exhibited within binary sequences ~\citep{Raff_Nicholas_2020}. Because ML algorithms can not usually handle more than a few thousand tokens of sequence length, the field has relied heavily on manually designed hash functions~\citep{Botacin_Galhardo, Breitinger_Astebol_Baier_Busch_2013, Lillis_Breitinger_Scanlon_2017, Oliver_Cheng_Chen_2013, Raff_Nicholas_2018, Roussev_2009, Winter_Schneider_Yannikos_2013}. 
In this work we will push deep learning-based sequence modeling to over 100,000 tokens, and longer sequences will be truncated down. Though this does not yet reach the full possible sequence length, it serves as a real-world task to determine the efficacy of our methods.

\subsection{Efficient Transformer-Based Models}
The quadratic cost of attention has motivated substantial research into more efficient architectures that maintain the performance of transformers. For smaller-scale models, there are a wide variety of such architectures \citep{performers,linear-trans,luna,linformer,bigbird}, however they are limited by their inability to be scaled and match the performance of traditional transformers.

Another approach to the quadratic run-time of attention that's gained popularity lately has been to simply pay it. Newer kernels for attention are reasonably fast in practice \citep{dao2023flashattention,dao2022flashattention} and new techniques for extending context length during post-training \citep{chen2023extending,peng2023yarn,roziere2023code}. However the expense of such models is impractical for many applications, as while they substantially decrease the costs associated with training long-context models they do not substantially decrease the memory overhead at inference time. This is essential because for most applications the primary bottleneck is GPU VRAM and not raw computing power.

\subsection{Non-Transformer Models for Sequences}
Recent research has also raised the prospect of alternatives to the transformer architecture for sequence-based tasks. Foremost among these are \textit{state-space models}, S4 \citep{s4} and its variants \citep{hippo, sgconv, poli2023hyena} which have achieved impressive performance on language and vision tasks. Previous work in malware detection has independently developed larger-width convolutions on the order of 128-256 wide kernels, followed by temporal pooling ~\citep{Raff2021,raff2017malware}

Simultaneously with this work, non-transformer architectures with more efficient inference-time context length scaling have begun to match the performance of transformers on natural language tasks\citep{gu2023mamba,peng2023rwkv} and pose an interesting area of exploration for future work in malware detection and other long-sequence problems.

\subsection{Our Contributions}
Our primary contributions are as follows:
\begin{enumerate}
    \item We introduce HGConv, a novel fusion of previous architectures \citep{sgconv,hrr} that achieves state-of-the-art performance on three standard malware classification benchmarks, and furthermore achieves its excellent performance with lower inter-run variance.
    \item We introduce novel algorithmic optimizations that enable HGConv to run substantially faster and with lower memory overhead than other global convolutional models.
    \item We show that the widely used Long Range Arena (LRA) \citep{lra} benchmark is a poor proxy for performance at malware classification, despite the fact that it is a task that requires reasoning about long contexts. This underlines the need for using domain-specific benchmarks whose construct validity has been validated in the real world instead of ``general performance'' benchmarks.
\end{enumerate}

\section{Methodology} 
In convolution, inputs are convolved with kernels or filters. Recent works have demonstrated the potential of global convolution in sequence modeling yet intricate kernel computation requires custom CUDA extensions to run \citep{s4} or crafted kernel design trying to make an approximation of the S4 kernel for each task \citep{sgconv}. In this paper, we focus on building a neuro-symbolic mechanism where kernels are defined as parameters and learned through auto-differentiation eliminating the necessity of intricate and detailed computations and task-specific kernel design. Before going over the details of the proposed HGConv, first we will give a brief overview of the HRR, and its properties, then the proposed method will be elaborated and finally, the algorithmic complexity will be delineated. Our implementation can be found at \url{https://github.com/FutureComputing4AI/HGConv}.

\subsection{Holographic Reduced Representations}
Holographic Reduced Representations (HRR) is a type of vector symbolic architecture (VSA) that represents compositional structure using circular convolution in distributed representations \citep{hrr}. In HRR, vector representations of properties and values can be combined together using circular convolution, and has been successfully used in recent literature~\citep{alam2023generalization,pmlr-v162-alam22a,menet2023mimo}. 

For instance, the color and shape of a red circle can be stored in a compressed representation using \emph{binding} operation ($\bind$) and additive properties of HRR by simply $\mathbf{b} = color \bind red + shape \bind circle$.  Here the abstract concepts ``color'', ``red'', ``shape'', and ``circle'' are arbitrarily assigned to a $d$ dimensional vector.  The method of retrieving knowledge from this compressed representation is known as \emph{unbinding} which is similar to binding operation with the inverse of a vector representation. Given vectors $x_i$, $y_i$ of dimension $d$, the binding operation is defined in \autoref{eq:binding}. 
\begin{equation} \label{eq:binding}
\mathbb{B} = x_i \bind y_i = \mathcal{F}^{-1}(\mathcal{F}(x_i) \odot \mathcal{F}(y_i))
\end{equation}
Here, $\mathcal{F}(\cdot)$ and $\mathcal{F}^{-1}(\cdot)$ refer to Fast Fourier Transform (FFT) and its inverse, respectively. To retrieve $x_i$ component from bound representation $\mathbb{B}$, the same binding operation is performed with the inverse of the $y_i$ vector component defined in \autoref{eq:inverse}. 
\begin{equation} \label{eq:inverse}
y_i^\dagger = \mathcal{F}^{-1}(\frac{1}{\mathcal{F}(y_i)})
\end{equation}
To extract the shape of the object in our example from $\mathbf{b}$, the unbinding operation is performed as $\mathbf{b} \bind shape^\dagger \approx circle$. Similarly, the same concept can be utilized to encode features by binding and decode by unbinding.

\begin{figure*}[!t]
\centerline{\includegraphics[width=0.8\textwidth]{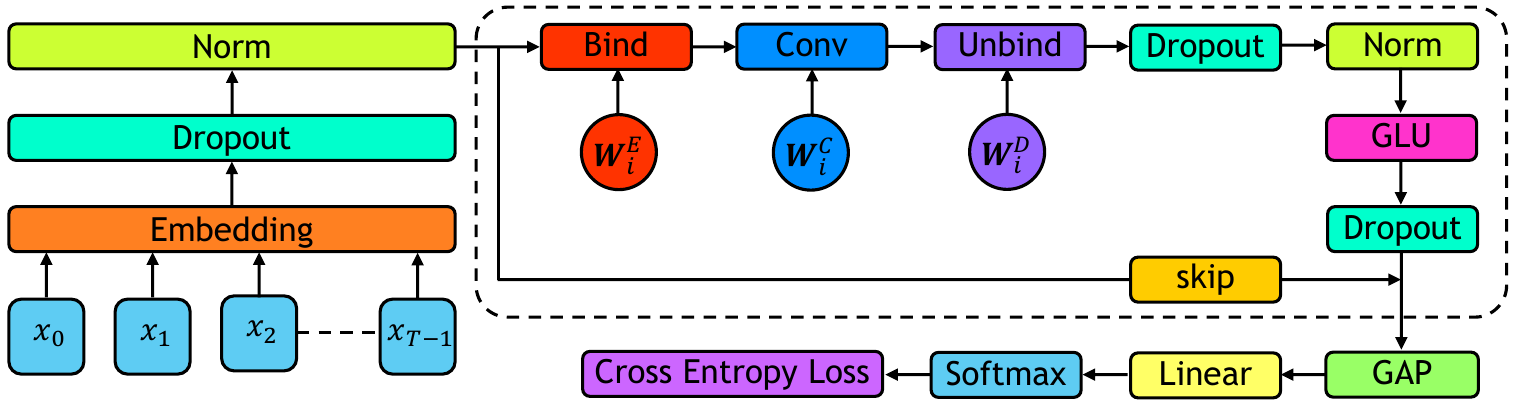}}
\caption{The block diagram of the proposed method. The dotted region shows a single layer of the proposed network which is repeated $N$ times. In the figure, \emph{prenorm} is applied. In the case of \emph{postnorm}, normalization is applied after the GLU layer before the skip connection.} 
\label{fig:block}
\end{figure*}

\subsection{Holographic Global Convolutional Networks}
We will learn both sequence-wise and depth-wise by integrating binding, global circular convolution, and unbinding operations subsequently. The filters for all the operations will be defined as parameters. First, the binding will be applied along the features which will encode kernel features with input features. Next, a global convolution will be applied along the elements of the sequence which will inter-mix the features of each sequence element. Finally, unbinding will decode the necessary useful features for learning. 
Since the binding step encodes the filter features to the input features, it will be denoted as the Encoder (E). For conciseness, circular convolution will be referred to as Conv or Convolution (C) unless otherwise specified, and likewise, the unbinding step will be deemed the Decoder (D).

Given an input sequence of $i$-th layer $\mathbf{X}_i \in \mathbb{R}^{T \times H}$ has $T$ tokens each having $H$ dimensional features, we will define three filter weights $\mathbf{W}_i^E \in \mathbb{R}^H, \mathbf{W}_i^C \in \mathbb{R}^{K \times H}, \mathbf{W}_i^D \in \mathbb{R}^H$ for encoder, convolution, and decoder, respectively where $K$ is the kernel dimension and $K \leq T$. $\mathbf{W}_i^C $ will be padded by zero up to maximum sequence length $T$ to perform global convolution.

The input features are encoded with encoder filter $\mathbf{W}_i^E$ using binding given in \autoref{eq:encoder1} and \autoref{eq:encoder2} \footnote{
\texttt{Notations:} \\ 
$\bind \;\: \rightarrow \texttt{binding ops}$ \\ 
$\circledast~\rightarrow \texttt{circular convolution}$ \\ 
$\odot~\rightarrow \texttt{elementwise multiplication}$ 
}. Here, each $m$-th element of feature vector $y_n$ of $ \mathbf{Y}_i$ is a linear combination of features where $\forall \ n, m \in \mathbb{N} : 0 \le n \le T-1, 0 \le m \le H-1$. The encoder step does not mix or alter the sequence elements. The sole attention is put on feature learning.
\begin{gather}
\mathbf{Y}_i = \mathbf{X}_i \bind \mathbf{W}_i^E \in \mathbb{R}^{T \times H} \label{eq:encoder1} \\ 
y_n[m] = \sum_{j=0}^{H-1} x_n[j] \ w_n^e[((m-j))_H] \label{eq:encoder2}
\end{gather}
After learning the features, the encoded features of each element are mixed with weighted input features, i.e., kernel $\mathbf{W}_i^C$ using convolution given in \autoref{eq:conv1}. Each feature vector $h[n]$ of the convolution layer is a linear weighted combination of encoded features of the tokens expressed in \autoref{eq:conv2} and \autoref{eq:conv3}. To include a bias term, a weight $\mathbf{W}_i^B \in \mathbb{R}^H$ is defined which is elementwise multiplied with $\mathbf{Y}_i$ is added, and consecutively a \textit{gelu} \citep{gelu} is applied.
\begin{gather} 
\mathbf{H}_i = \mathbf{Y}_i \circledast \mathbf{W}_i^C + \mathbf{Y}_i \otimes \mathbf{W}_i^B \in \mathbb{R}^{T \times H} \label{eq:conv1} \\ 
\mathbf{Y}_i \circledast \mathbf{W}_i^C \ \textbf{:} \ h[n] = \sum_{j=0}^{T-1} y[j] \ w^c[((n-j))_T] \label{eq:conv2} \\ 
\begin{split}
h[n] = y_0 w_n^c + y_1 w_{n-1}^c + \cdots + y_n w_0^c + \\ 
y_{n+1} w_{T-1}^c + y_{n+2} w_{T-2}^c + \cdots + y_{T-1} w_{n+1}^c \label{eq:conv3}
\end{split}
\end{gather}
Since unbinding can extract information from the added feature vectors, it will be utilized to decode useful features from the convolutional step. Given that features are mixed regardless of their significance, by learning appropriate kernels, the most important features can be extracted using unbinding. Specifically, the unbinding step is expected to learn to get rid of overmixed or unnecessarily mixed element features.

\begin{gather} \label{eq:decoder}
\mathbf{Z}_i = \mathbf{H}_i \bind {\mathbf{W}_i^D}^\dagger \in \mathbb{R}^{T \times H} \\ 
z_n[m] = \sum_{j=0}^{H-1} h_n[j] \ w_n^d[((m-j))_H]
\end{gather}

The extracted features are processed by a gated linear unit (GLU) \citep{glu} given in \autoref{eq:glu} and subsequently a dropout layer is used. $\mathbf{W}_i^\alpha$ and $\mathbf{W}_i^\beta$ are the weights are GLU unit and $\sigma$ is the sigmoid activation.

\begin{equation} \label{eq:glu}
\mathbf{G}_i = \mathbf{W}_i^\alpha \ \mathbf{Z}_i \ \odot \ \sigma(\mathbf{W}_i^\beta \ \mathbf{Z}_i) 
\end{equation}

Finally, a skip connection is used by adding the unperturbed input $\mathbf{X}_i$ to the processed feature from GLU unit $\mathbf{G}_i$. The output of the $i$-th layer $\mathbf{X}_{i+1}$ can be fed to the next layer the process can be repeated $N$ times to extract the deeper features by the combinations of \texttt{bind} $\rightarrow$ \texttt{conv} $\rightarrow$ \texttt{unbind} $\rightarrow$ \texttt{glu} units in each layer to improve the performance of the network.

\begin{equation}
\mathbf{X}_{i+1} = \mathbf{G}_i + \mathbf{X}_{i}
\end{equation}

A generic block diagram of the proposed method is presented in \autoref{fig:block}. In the embedding layer, both word and position embeddings are used and added together. For normalized floating point inputs, a linear layer is used in place of word embedding. In the norm layer, either layer normalization \citep{ln} or batch normalization \citep{bn} can be employed. The global average pooling (GAP) is applied to the output of the $N$-th layer which is subsequently fed to a linear layer with a feature size the same as the number of classes. The loss is calculated using the softmax cross-entropy loss function which is optimized using the Adam optimizer where a cosine decay learning rate scheduler with warmup is employed.

\subsection{Algorithmic Complexity}
The time complexity of the main three layers, i.e., binding, convolution, and unbinding are $\mathcal{O}(T \cdot H\log{H})$, $\mathcal{O}(T \log{T} \cdot H)$, and $\mathcal{O}(T \cdot H\log{H})$, respectively. Therefore, the overall time complexity is $\mathcal{O}(T \log{T})$ log-linear with respect to the sequence length $T$. Since in all the layers, the shape of the tensors is $T \times H$, the space complexity is $\mathcal{O}(T)$ linear. Feature dimension $H$ is assumed to be constant. A step-by-step breakdown of the time and space complexity is given in \autoref{eq:time} and \autoref{eq:space}.
\begin{equation} \label{eq:time}
\begin{split}
 &   \textsc{Time Complexity} \\ 
 & = \mathcal{O}(T \cdot H\log{H} + T \log{T} \cdot H + T \cdot H\log{H}) \\ 
 & = \mathcal{O}(2 \times T \cdot H\log{H} + T \log{T} \cdot H) \\ 
 & = \mathcal{O}(T + T \log{T}) \texttt{ [H is constant]} \\  
 & = \mathcal{O}(T \cdot \{1 + \log{T}\}) \\ 
 & = \mathcal{O}(T \log{T}) \texttt{ log-linear} \\ 
\end{split}
\end{equation}
\begin{equation} \label{eq:space}
\begin{split}
 & \textsc{Space Complexity} \\ 
 & = \mathcal{O}(T \cdot H) \\ 
 & = \mathcal{O}(T) \texttt{ linear [H is constant]} \;\;\;\;\;\;\;\;\;\;\;\;\;\;\; \\ 
\end{split}
\end{equation}

\section{Experiments and Results}
In this paper, we are proposing a neuro-symbolic method of sequence processing that encode feature, convolve along all the sequence elements, and finally decode necessary features compensating for overmixing. To validate the proposed method, experiments are performed focusing on practical applications where long sequences are a common phenomenon such as malware classification where sequence length can reach up to $\approx$ 200M. In our experiments, we will adopt well-known malware classification benchmarks such as the Microsoft Windows Malware benchmark that comes from the 2015 Kaggle competition \citep{kaggle}, Android application packages (APK) Malware benchmark from Drebin dataset \citep{drebin}, and EMBER malware classification benchmark \citep{ember}. As will be seen in the results, in most cases existing hash-based algorithms that have no learning phase outperform existing Transformer and similar long-sequence learning algorithms.

\paragraph{Kaggle}
Microsoft Malware Classification Challenge (BIG 2015) hosted on Kaggle \citep{kaggle} is a benchmark of 9 Windows malware families. The dataset contains $10,868$ samples total uncompressed size of $184$ GB which is split into train and test set by $80-20$ ratio per class by random sampling. Each of the data samples comes in two different forms, in one form it is the raw binary of the original executables referred to as \textsc{Kaggle Raw} of size $47$ GB, and in another form, it is the human-readable assembly referred to as \textsc{Kaggle Asm} of size $137$ GB. \textsc{Asm} files are generated by IDA Pro which contains additional features that seem to make it easier to learn. However, it is also $\approx 3 \times$ larger with longer sequence lengths than \textsc{Raw} files, thus, balancing the difficulty of the dataset.

\paragraph{Drebin}
Android APK namely Drebin \citep{drebin} is a benchmark of $178$ malware families containing $5,560$ samples total uncompressed size of $16$ GB. Nevertheless, $70\%$ of the families contains less than 10 samples and $88.8\%$ of the families contains less than 40 samples. Therefore, to be able to learn from enough data, in our experiments we have utilized top $20$ malware families containing  $4,664$ samples of size $14$ GB which is split into train and test set by $80-20$ ratio per class. The original data of the dataset is in APK format which is referred to as \textsc{Drebin Apk} of size $6$ GB. Like Kaggle, another version of the dataset is built by converting the APK files to uncompressed TAR files which have a size of $8$ GB and are referred to as \textsc{Drebin Tar}. Since the difference between the samples is the amount of compression, it will be useful to understand how compression is handled by each algorithm.

\paragraph{EMBER}
EMBER is binary malware classification benchmark \citep{ember} containing $800K$ samples of Windows executable files of total $1.02$ TB of size. Among them, the training split contains $300K$ benign and $300K$ malicious files of a total size of $826$ GB. On the other hand, the test split contains $100K$ benign and $100K$ malicious files of a total size of $220$ GB. Although the sequence length of the files in the EMBER dataset can be over $100M$ long which is not practical to process by any sequence model, we start our experiments with a relatively shorter length of $256~(2^8)$ which is exponentially incremented until $131,072~(2^{17})$. Since most of the important features are encoded at the beginning of the sequence, we could not see any benefit of using a much longer sequence length than $2^{17}$.

\subsection{Training} 
The sequences of inputs are padded or truncated up to the maximum sequence length to train the proposed HGConv network. To suppress the embedding for the padded tokens binary mask is produced and multiplied by the embedding matrix. In the convolutional step, the kernel dimension $K$ can be smaller than the actual sequence length which is also padded with zeros up to the maximum sequence length $T$ to perform FFT convolution. Since all the tasks are essentially classification, to train the network, the softmax cross-entropy loss function is employed which is optimized using the Adam optimizer with cosine scheduler learning rate. Moreover, label smoothing is applied with a smoothing factor $\alpha=0.1$. The hyperparameter used in each of the tasks is fine-tuned and optimized. The list of the hyperparameters used in each task is presented in \autoref{sec:appendix_b}. The training is performed on a single node 16 NVIDIA TESLA PH402 32GB GPU machine where the mean of the gradient from each machine is used to update the parameters.

\begin{figure*}[!htbp]
\centerline{\includegraphics[width=\textwidth]{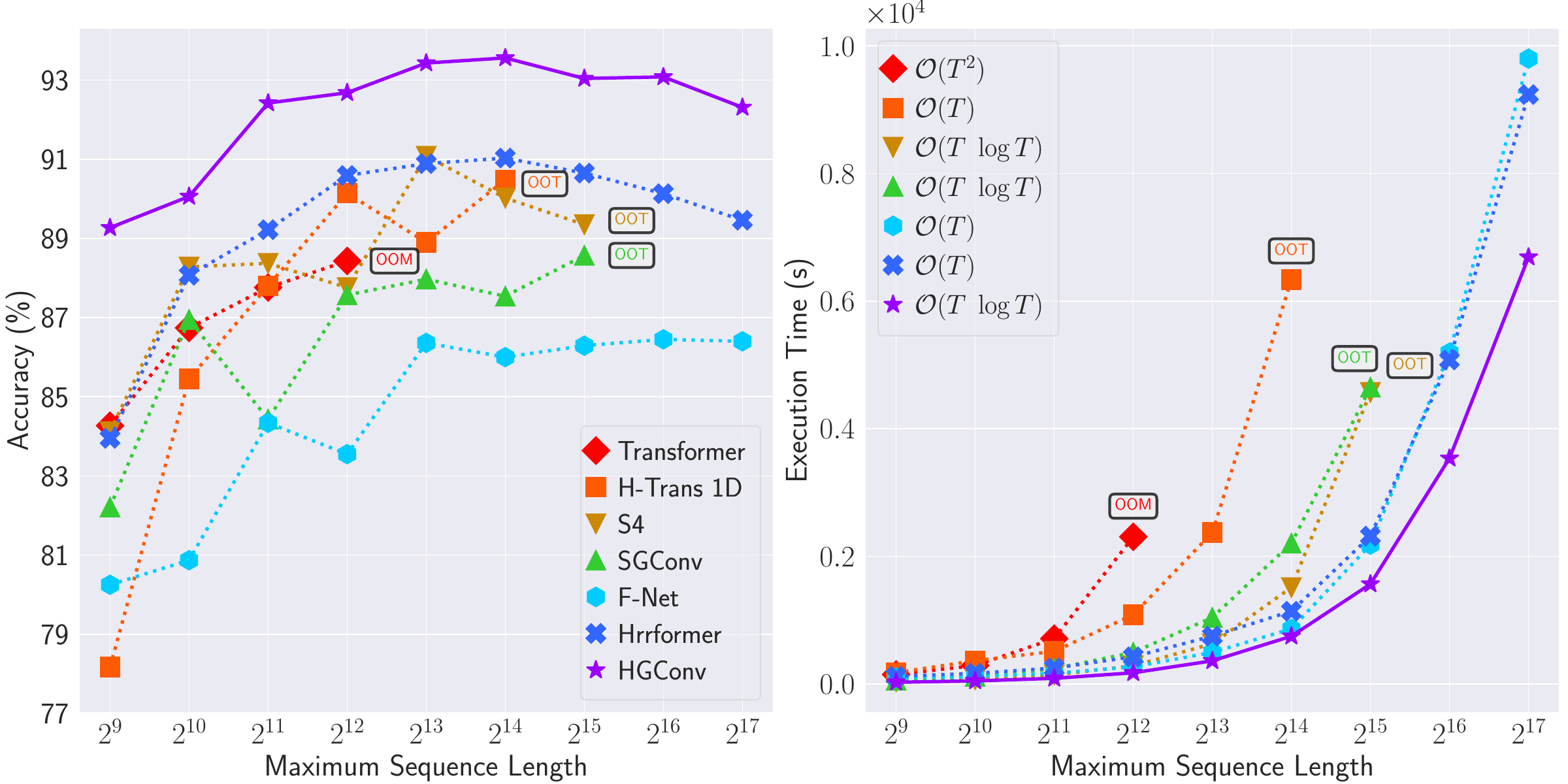}}
\caption{Ember long sequence malware classification results. In the figure, OOT and OOM stand for out-of-time and memory shown for models that face such issues after a particular sequence length. The figure shows a shorter comparison. A broader comparison with additional models Linformer \citep{linformer}, Performer \citep{performers}, and F-Net \citep{fnet} and numeric results are presented in \autoref{sec:appendix_d}.} 
\label{fig:ember_short}
\end{figure*}

\subsection{Evaluations}
\begin{figure*}[!htbp]
\centerline{\includegraphics[width=\textwidth]{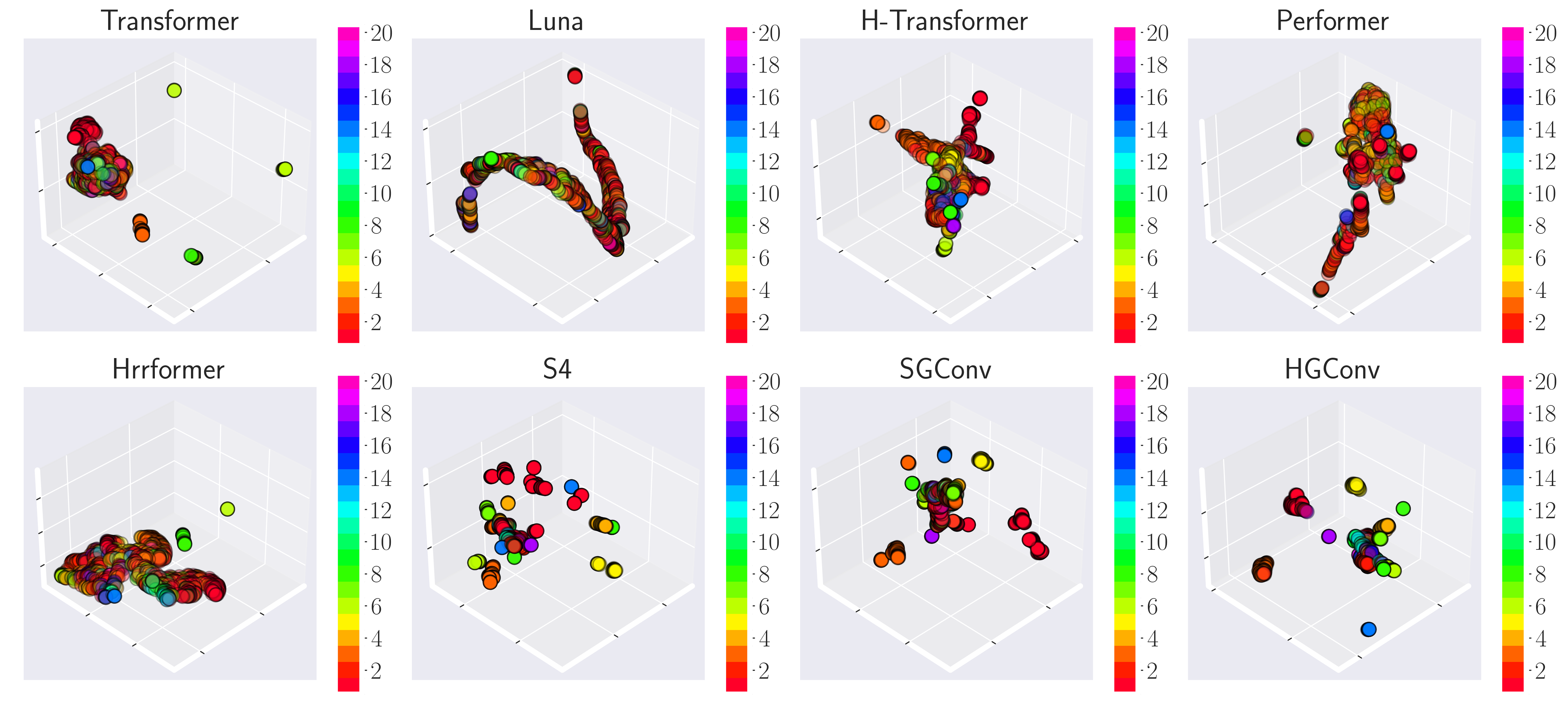}}
\caption{\textsc{Drebin Apk} dataset in the benchmark has the most variation in the results across the models. The figure shows the UMAP 3D representation of the output from the penultimate layer of all the models for \textsc{Drebin Apk}. The better the clusters the higher the accuracy.} 
\label{fig:drebin_umap}
\end{figure*}

To evaluate the performance, the proposed HGConv is compared with other state-of-the-art (SOTA) sequence models. For \textsc{Kaggle} and \textsc{Drebin} datasets, the proposed method is compared with non-attention-based processors such as Lempel-Ziv Jaccard Distance (LZJD)~\citep{pylzjd-proc-scipy-2019,RAFF201834,10.1145/3097983.3098111}, Stochastic Hashed Weighted Lempel-Ziv (SHWeL) \citep{raff2017malware}, attention-based processors such as Transformer \citep{transformer}, Performer \citep{performers}, Hrrformer \citep{hrrformer}, and state space model based processors S4 \citep{s4} and SGConv \citep{sgconv}. Other compression-based methods like Burrows-Wheeler Markov Distance (BWMD)~\citep{BWMD} and Lempel-Ziv Networks~\citep{pmlr-v187-saul23a} were not considered due to lower accuracy compared to the selected baselines, and their other benefits are not a focus of this work. \autoref{tab:kaggle_drebin} shows the mean accuracy with standard deviation for 10-fold cross-validation for each of the methods. Among all the methods, the proposed HGConv achieved the best results for all the datasets with the smallest standard deviation. It is also the only method to out-perform the existing hash-based approaches, showing how existing methods did not adequately learn from long sequence problems. In terms of fluctuation among the models, the variation in the results of \textsc{Drebin Apk} is the most noticeable. \autoref{fig:drebin_umap} shows the UMAP 3D representation \citep{umap,Nolet2021} of the output of the penultimate layer of all the models which reveals the clustering patterns. HGConv has visibly better clusters which makes the final layer classifier predict correctly. Moreover, qualitative inspection shows that models that perform better generally show clearer and better separated clusters, with HGConv in particular showing the best clustering behavior.

\begin{table*}[!t]
\centering
\caption{Results of 10-fold cross-validation on Kaggle Microsoft Malware Classification Challenge and Drebin Android Malware classification. Values inside the parenthesis are standard deviations. Also, for both of the datasets, the training time per epoch is provided in seconds.}
\vspace{1mm}
\label{tab:kaggle_drebin}
\renewcommand{\arraystretch}{1.05}
\resizebox{\textwidth}{!}{
\begin{tabular}{lcccccc}\toprule

\multirow{2}{*}{\textsc{Model}} & \multirow{2}{*}{\textsc{\shortstack{Kaggle\\Raw}}} & \multirow{2}{*}{\textsc{\shortstack{Kaggle\\Asm}}} & \multirow{2}{*}{\textsc{\shortstack{Kaggle\\Time}}} & \multirow{2}{*}{\textsc{\shortstack{Drebin\\Apk}}} & \multirow{2}{*}{\textsc{\shortstack{Drebin\\Tar}}} & \multirow{2}{*}{\textsc{\shortstack{Drebin\\Time}}} \\
 &  &  &  &  &  &  \\ \midrule

LZJD \citep{10.1145/3097983.3098111}      & 97.6 (1.50) & 97.1 (6.10) & -- & 80.8 (2.60) & 81.0 (6.50) & -- \\ 
1NN-SHWeL \citep{raff2017malware} & 97.6 (1.38) & 97.3 (1.93) & -- & 83.6 (1.94) & 87.9 (1.84) & -- \\ 
LR-SHWeL \citep{raff2017malware}  & 96.7 (2.07) & 96.9 (2.08) & -- & 78.4 (2.26) & 89.1 (2.29) & -- \\ \midrule

Transformer \citep{transformer} & 72.68 (3.77) & 95.60 (1.52) & 31.55 & 40.13 (6.11) & 69.50 (2.67) & 15.90 \\
F-Net \citep{fnet}              & 93.17 (1.08) & 95.74 (1.03) & 6.54  & 69.41 (1.81) & 80.98 (1.22) & 4.73 \\
Luna-256 \citep{luna}           & 89.50 (0.89) & 93.47 (0.96) & 26.19 & 24.30 (2.36) & 56.42 (7.90) & 16.49 \\ 
H-Transformer \citep{h-transformer} & 92.78 (0.49) & 98.07 (0.29) & 117.53 & 71.85 (0.84) & 87.40 (0.70) & 99.64 \\
Performer \citep{performers}      & 94.63 (0.79) & 97.66 (0.48) & 37.08 & 70.44 (3.65) & 82.38 (1.12) & 18.31 \\ 
Hrrformer \citep{hrrformer}       & 94.41 (0.57) & 98.52 (0.23) & 7.35 & 57.28 (3.80) & 84.07 (1.03) & 5.42 \\
\midrule
S4 \citep{s4}                     & 96.44 (0.41) & 98.66 (0.32) & 17.51 & 88.38 (1.69) & 87.94 (1.05) & 14.97 \\ 
SGConv \citep{sgconv}             & 95.13 (0.91) & 98.12 (0.56) & 24.37 & 76.23 (3.14) & 80.04 (4.33) & 24.37 \\
\midrule
\textbf{HGConv}  & \textbf{98.86 (0.12)} & \textbf{99.63 (0.14)} & \textbf{5.86} & \textbf{90.15 (0.47)} & \textbf{91.86 (0.35)} & \textbf{3.63} \\ \bottomrule
\end{tabular}
}
\end{table*}

EMBER is a benchmark with very long sequences. In our experiments, we started with a moderate sequence length of $256~(2^8)$ and incremented up to $131,072~(2^{17})$ and computed the accuracy and execution time for each sequence length which is presented in \autoref{fig:ember_short}. For practical reasons, we have set a maximum limit of $10,000$ seconds per epoch. If a method takes more than that is marked as out-of-time (OOT). If the model and data can not be put onto the memory for a particular sequence length that is marked as out-of-memory (OOM) in the figure. HGConv not only achieves the best accuracy but also takes the least amount of time among all the compared methods. The full comparison and all the numerical results are presented in \autoref{sec:appendix_d}. We also find that HGConv runs substantially faster than all other methods, achieving far more efficient scaling despite the increased theoretical complexity compared to Hrrformer and F-Net.

\begin{table*}[!t]
\centering
\caption{LRA benchmark scores. HGConv is from this work, while Hrrformer, S4, and SGConv scores are from their respective papers. All other scores are from \citep{lra}. \citep{lra} and \citep{hrrformer} report that models ``do not learn anything'' on Path-X, shown here with a \xmark. We observe this happening with HGConv as well.}
\renewcommand{\arraystretch}{1.1}
\adjustbox{max width=\textwidth}{
\begin{tabular}{@{}lccccccc@{}}
\toprule
\textsc{Model} & \textsc{ListOps} & \textsc{Text} & \textsc{Retrieval} & \textsc{Image} & \textsc{Pathfinder} & \textsc{Path-X} & \textsc{Average} \\ \midrule
Random & 10.00 & 50.00 & 50.00 & 10.00 & 50.00 & 50.00 & 36.67 \\ \midrule
Transformer \citep{transformer}  & 36.37 & 64.27 & 57.46 & 42.44 & 71.40 & \xmark & 54.39 \\
Linformer \citep{linformer}      & 35.70 & 53.94 & 52.27 & 38.56 & 76.34 & \xmark & 51.36 \\
Performer \citep{performers}     & 18.01 & 65.40 & 53.82 & 42.77 & 77.05 & \xmark & 51.41 \\
F-Net \citep{fnet}                & 35.33 & 65.11 & 59.61 & 38.67 & 77.80 & \xmark & 55.30 \\
Luna-256 \citep{luna}            & 37.25 & 64.57 & 79.29 & 47.38 & 77.72 & \xmark & 61.24 \\ 
H-Transformer \citep{h-transformer} & 49.53 & 78.69 & 63.99 & 46.05 & 68.78 & \xmark & 61.41 \\
Hrrformer \citep{hrrformer}      & 39.98 & 65.38 & 76.15 & 50.45 & 72.17 & \xmark & 60.83 \\\midrule
S4 \citep{s4} & \underline{59.60} & 86.82 & \underline{90.90} & \textbf{88.65} & \underline{94.20} & \underline{96.35} & \underline{86.09} \\
SGConv \citep{sgconv} & \textbf{61.45} & \textbf{89.20} & \textbf{91.11} & \underline{87.97} & \textbf{95.46} & \textbf{97.83} & \textbf{87.17} \\ 
\midrule
\textbf{HGConv} & 49.75 & \underline{88.15} & 90.62 & 85.08 & 92.04 & \xmark & 81.13 \\
\bottomrule
\end{tabular}
}
\label{tab:lra}
\end{table*}

\begin{table*}[!t]
\centering
\caption{Rank order performance of models on each benchmark. For \textsc{Ember} we use sequences of length up to $2^{14}$ as that's the maximum size found in the LRA Benchmark.}
\renewcommand{\arraystretch}{1.05}
\begin{tabular}{lcccccc}
\toprule
\textsc{Model} & \textsc{K. Raw}  & \textsc{K. Asm} & \textsc{D. Apk} & \textsc{D. Tar} & \textsc{Ember} ($2^{14}$) & \textsc{LRA}\\ \midrule
Transformer \citep{transformer}    & 9 & 8 & 8 & 8 & 9 & 8 \\
Performer \citep{performers}       & 3 & 7 & 4 & 5 & 7 & 9 \\ 
F-Net \citep{fnet}                 & 3 & 6 & 6 & 7 & 5 & 7 \\
Luna-256 \citep{luna}              & 8 & 9 & 9 & 9 & 8 & 5 \\ 
H-Transformer \citep{h-transformer}& 7 & 4 & 4 & 2 & 3 & 4 \\
Hrrformer \citep{hrrformer}        & 6 & 2 & 7 & 4 & 2 & 6 \\ \midrule
S4 \citep{s4}                      & 2 & 2 & 2 & 2 & 4 & 2 \\ 
SGConv \citep{sgconv}              & 3 & 4 & 3 & 6 & 5 & \textbf{1} \\
\midrule
\textbf{HGConv} (ours)  & \textbf{1} & \textbf{1} & \textbf{1} & \textbf{1} & \textbf{1} & 3 \\ \bottomrule
\end{tabular}%
\label{tab:rank}
\end{table*}

\section{Long Range Arena Does Not Predict EMBER Reliably}
Recent work on benchmarking large language models \citep{eval-harness,raji2021ai} has questioned the construct validity of the widespread practice of assuming that ``diverse'' acontextual benchmarks are indicative of performance on tasks of interest. For long-context models, this is exemplified by the widespread use of the Long Range Arena \citep{lra}, which contains tasks that evaluate parsing long expressions, classifying movie reviews, assessing text similarity, classifying flattened CIFAR-10 images, and identifying if two points are connected by a long path. Despite the lack of relevance of these tasks to their application domains, LRA scores have been used to motivate architectural design choices in work in genomics \citep{nguyen2023hyenadna,romero2023dnarch}, analyzing ECGs \citep{zama2023ecg}, speech enhancement \citep{du2023spiking}, and reinforcement learning \citep{lu2023structured}.

Long Range Arena (LRA) is a benchmark of 6 tasks covering diverse problem areas with different modalities. The \textsc{ListOps} task deals with the hierarchically structured data of mathematical operations with delimiters with $96K$ training and $2K$ test data. In \textsc{Text} task, IMDB movie review \citep{imdb} text dataset is employed. Classification is performed character level to include additional complexity. The task has a balanced train-test split of size $25K$. The \textsc{Retrieval} task models the textual similarity of two documents for which the ACL Anthology Network (AAN) \citep{aan} dataset is utilized with $147K$ training and $17K$ test samples. \textsc{Image} task comprises of grayscale sequential CIFAR-10 image classification that puts the hurdle of 2D spatial relations into a 1D sequence of pixels. Finally, the \textsc{Pathfinder} and \textsc{Path-X} are the binary classification tasks containing grayscale images of dotted lines and circles connected or disconnected introduced in \citep{pathfinder}. The difference between them is the sequence length from  $1K$ to $16K$ both containing $160K$ training and $20K$ test samples.

\begin{figure}[!htbp]
\centerline{\includegraphics[width=\columnwidth]{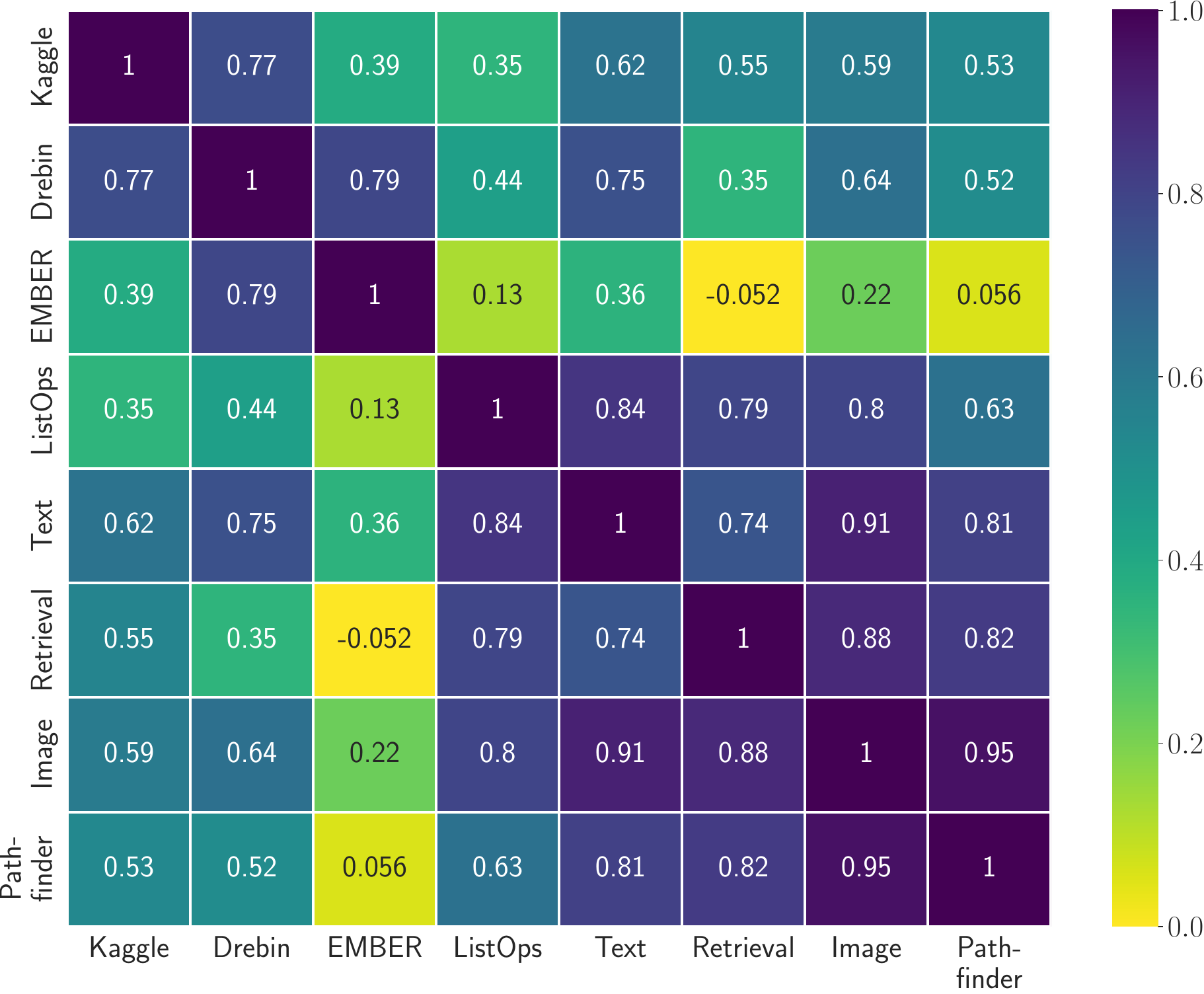}}
\caption{The correlation between Malware and other LRA tasks accuracies. While performance between LRA tasks is highly correlated with one another, they all correlate far worse with the malware benchmarks.} 
\label{fig:correlation}
\end{figure}

We investigate the Long Range Arena and find that average performance is uncorrelated with performance on any of our malware tasks. While performance between LRA tasks is highly correlated with one another, they all correlate far worse with performance on malware task benchmarks shown in \autoref{fig:correlation}.

In terms of performance, in text classification, HGConv achieved the second-best score of $88.15\%$ and overall third-based average accuracy of $81.13\%$ in the LRA benchmark presented in \autoref{tab:lra}. When comparing the rank order of model performance on LRA to our malware tasks, we see that LRA scores are not very predictive of performance on the malware benchmarks as shown in \autoref{tab:rank}. LRA rates S4 and SGConv well ahead of the other models, while their performance is far less outstanding on malware benchmarks. SGConv in particular has a median ranking of fourth on our malware benchmarks, despite being the clear best model on LRA. The actual best malware model, HGConv, only beats S4 or SGConv on one of the six LRA tasks.

\section{Conclusion}
In this paper, we have introduced a long-range global convolutional network named HGConv by utilizing the properties of vector symbolic architecture called HRR. The proposed network facilitates learning by encoding and decoding features. Our network does not require a custom CUDA extension to run or have any intricate kernel design, unlike other existing global convolutional networks. Rather kernel is defined as parameters and learnt through auto-differentiation. In this work, we particularly focused on a real-world application of long-range models in malware detection. On Kaggle and Drebin benchmarks, the proposed method scored a new SOTA of $99.3\%$ and $91.0\%$, respectively. 

Results on EMBER demonstrate both superior accuracy and execution time where experiments are performed for various sequence lengths. HGConv has not only achieved a new highest accuracy of $93.56\%$ for sequence length $16,384$ but also consistently consumed the least amount of time to execute. We have also investigated the performance and rank order of LRA and found that they all correlate far worse with malware performance and do not predict EMBER reliably. In conclusion, the experimental results, and comparison demonstrate the fidelity of the proposed method in a practical application like malware detection.

\section*{Acknowledgement}

We thank Maya Fuchs for her feedback and copy-editing on this paper.

\bibliography{refs}

\section*{Checklist}

\begin{enumerate}

\item For all models and algorithms presented, check if you include:
\begin{enumerate}
\item A clear description of the mathematical setting, assumptions, algorithm, and/or model. [Yes]
\item An analysis of the properties and complexity (time, space, sample size) of any algorithm. [Yes]
\item (Optional) Anonymized source code, with specification of all dependencies, including external libraries. [Yes/No/Not Applicable]
\end{enumerate}

\item For any theoretical claim, check if you include:
\begin{enumerate}
\item Statements of the full set of assumptions of all theoretical results. [Yes]
\item Complete proofs of all theoretical results. [Yes]
\item Clear explanations of any assumptions. [Yes]     
\end{enumerate}

\item For all figures and tables that present empirical results, check if you include:
\begin{enumerate}
\item The code, data, and instructions needed to reproduce the main experimental results (either in the supplemental material or as a URL). [Yes]
\item All the training details (e.g., data splits, hyperparameters, how they were chosen). [Yes]
     \item A clear definition of the specific measure or statistics and error bars (e.g., with respect to the random seed after running experiments multiple times). [Yes]
     \item A description of the computing infrastructure used. (e.g., type of GPUs, internal cluster, or cloud provider). [Yes]
\end{enumerate}

\item If you are using existing assets (e.g., code, data, models) or curating/releasing new assets, check if you include:
\begin{enumerate}
\item Citations of the creator If your work uses existing assets. [Yes]
\item The license information of the assets, if applicable. [Yes]
\item New assets either in the supplemental material or as a URL, if applicable. [Yes]
\item Information about consent from data providers/curators. [Yes]
\item Discussion of sensible content if applicable, e.g., personally identifiable information or offensive content. [Not Applicable]
\end{enumerate}

\item If you used crowdsourcing or conducted research with human subjects, check if you include:
\begin{enumerate}
\item The full text of instructions given to participants and screenshots. [Not Applicable]
\item Descriptions of potential participant risks, with links to Institutional Review Board (IRB) approvals if applicable. [Not Applicable]
\item The estimated hourly wage paid to participants and the total amount spent on participant compensation. [Not Applicable]
\end{enumerate}

\end{enumerate}

\newpage
\appendix
\onecolumn

\section{Hyperparameters} \label{sec:appendix_b}
The hyperparameters used in the experiments are presented in \autoref{tab:hyper}. For Kaggle, Drebin, and EMBER no weight decay is used. For all the LRA tasks, weight decay of $0.05$ is used except for Pathfinder where weight decay rate is $0.03$. %

\begin{table*}[!htbp]
\caption{The hyperparameters used in each of the experiments. LN and BN refer to Layer Norm and Batch Norm. When the pre-norm is True, no post-normalization is used, and vice versa. For EMBER, experiments are performed on a variable sequence length $T$ and the batch size is chosen according to the given expression to fit the data on memory.}
\label{tab:hyper}
\renewcommand{\arraystretch}{1.2}
\adjustbox{max width=\columnwidth}{
\begin{tabular}{@{}lccccccccccc@{}}
\toprule
\multirow{2}{*}{} & \multirow{2}{*}{\textsc{Norm}} & \multirow{2}{*}{\textsc{\shortstack{Pre\\Norm}}} & \multirow{2}{*}{\textsc{\shortstack{Batch\\Size}}} & \multirow{2}{*}{\textsc{\shortstack{Vocab\\Size}}} & \multirow{2}{*}{\textsc{\shortstack{Sequence\\Length}}} & \multirow{2}{*}{\textsc{\shortstack{Kernel\\Dim}}} & \multirow{2}{*}{\textsc{Features}} & \multirow{2}{*}{\textsc{Dropout}} & \multirow{2}{*}{\textsc{Layers}} & \multirow{2}{*}{\textsc{\shortstack{Learning\\Rate}}} & \multirow{2}{*}{\textsc{Epochs}} \\ 
 &  &  &  &  &  &  &  &  &  &  &  \\ \toprule
\textsc{Kaggle}     & LN & True & 64 & 257 & 4096 & 32 & 256 & 0.1 & 1 & 0.01 & 10 \\ 
\textsc{Drebin}     & LN & True & 32 & 257 & 4096 & 32 & 256 & 0.1 & 1 & 0.01 & 10 \\ 
\textsc{EMBER}      & LN & True & $\max(2^{16 - \log_{2}{T}}, 1)$ & 257 & $T$ & 32 & 256 & 0.1 & 1 & 0.01 & 10 \\ \toprule
\textsc{ListOps}    & BN & False & 100 & 17  & 2000 & 64  & 128 & 0.0 & 6 & 0.01 & 40 \\ 
\textsc{Text}       & BN & True  & 50  & 257 & 4096 & 7   & 512 & 0.0 & 2 & 0.01 & 32 \\ 
\textsc{Retrieval}  & LN & True  & 128 & 128 & 4000 & 4   & 128 & 0.0 & 6 & 0.01 & 30 \\ 
\textsc{Image}      & LN & True  & 50  & 256 & 1024 & 128 & 512 & 0.2 & 6 & 0.01 & 200 \\ 
\textsc{Pathfinder} & BN & False & 64  & 256 & 1024 & 128 & 512 & 0.0 & 4 & 0.004 & 200 \\ \bottomrule
\end{tabular}
}
\end{table*}

\section{UMAP 3D Representations} 
\begin{figure*}[!htbp]
\centerline{\includegraphics[width=\textwidth]{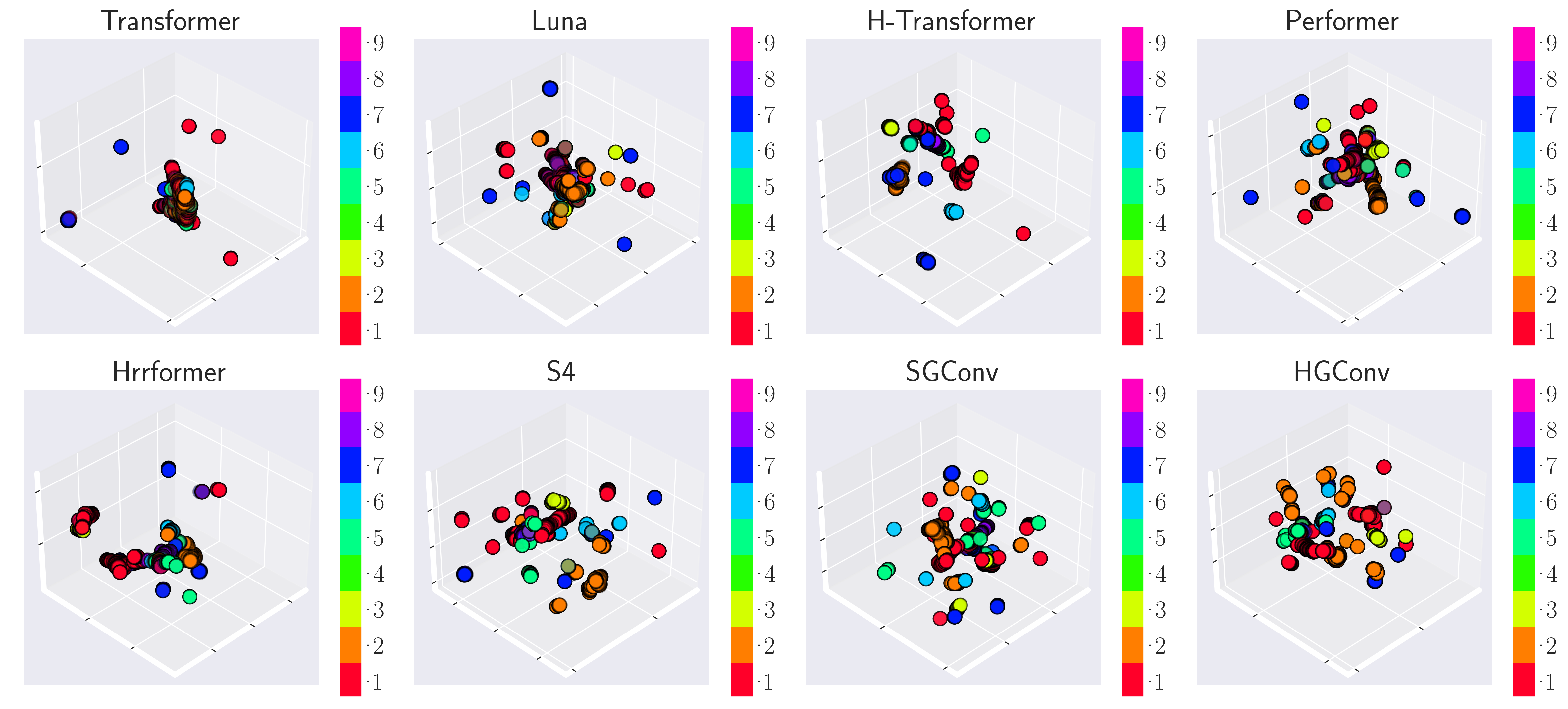}}
\caption{\textsc{Kaggle Raw}} 
\label{fig:kaggle_raw_umap}
\end{figure*}

\begin{figure*}[!htbp]
\centerline{\includegraphics[width=\textwidth]{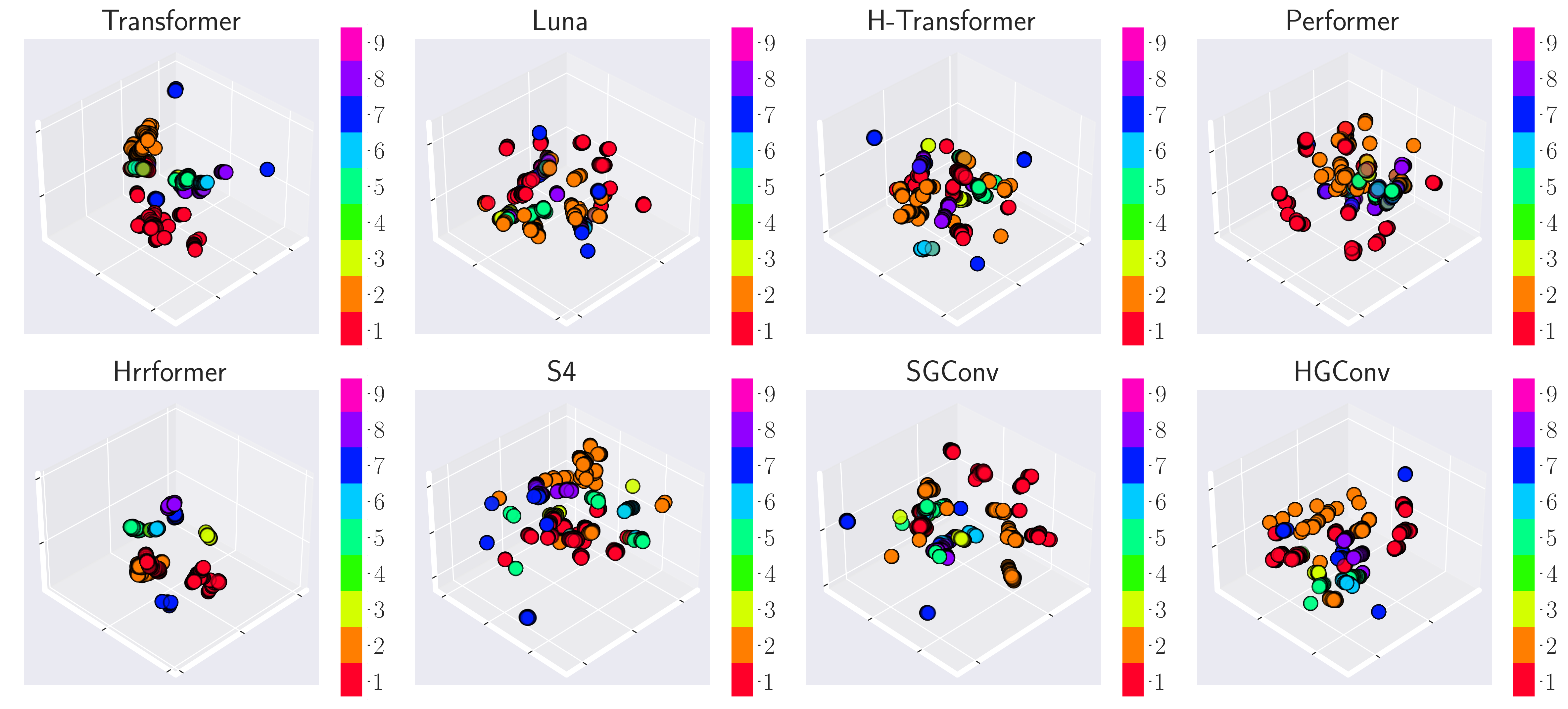}}
\caption{\textsc{Kaggle Asm}} 
\label{fig:kaggle_asm_umap}
\end{figure*}

\begin{figure*}[!t]
\centerline{\includegraphics[width=\textwidth]{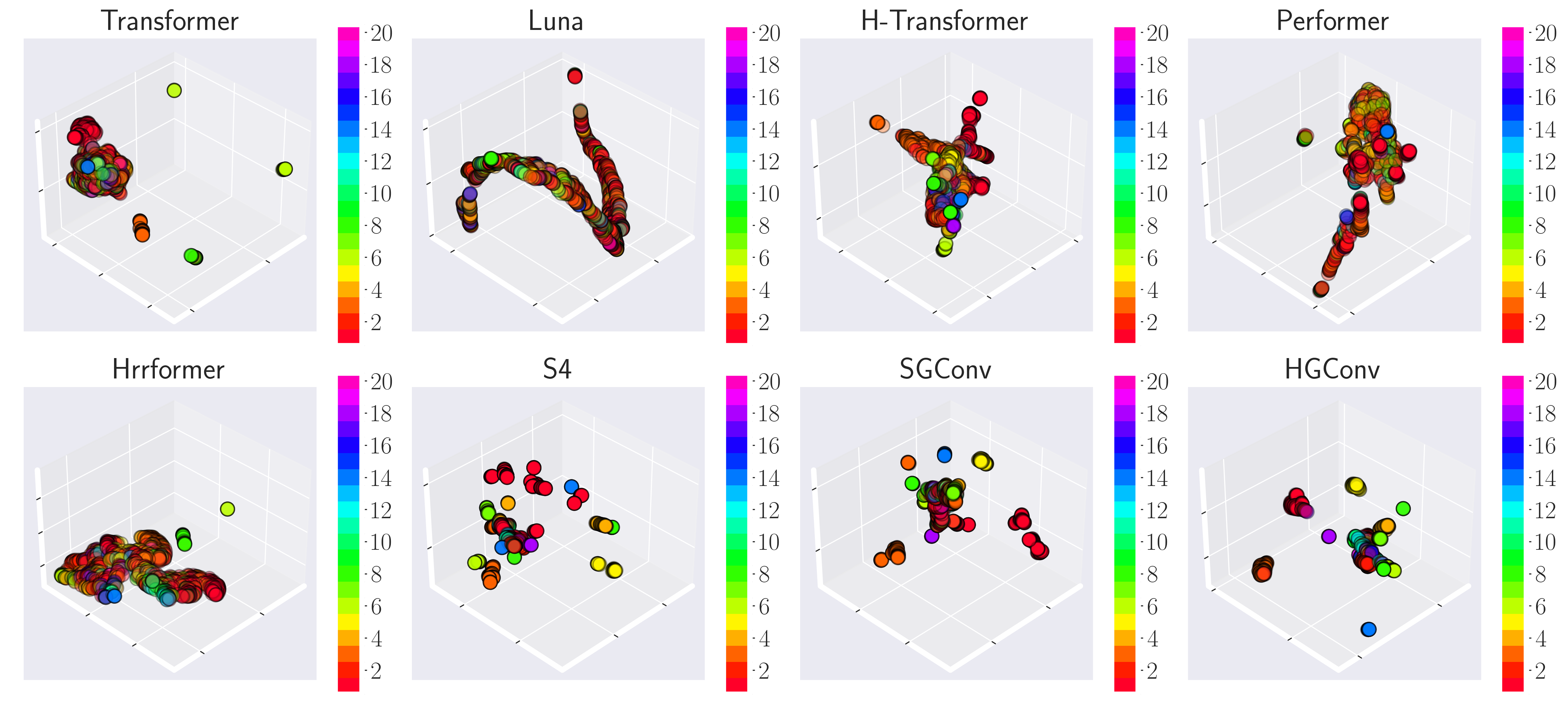}}
\caption{\textsc{Drebin Apk}} 
\label{fig:drebin_apk_umap}
\end{figure*}

\begin{figure*}[!t]
\centerline{\includegraphics[width=\textwidth]{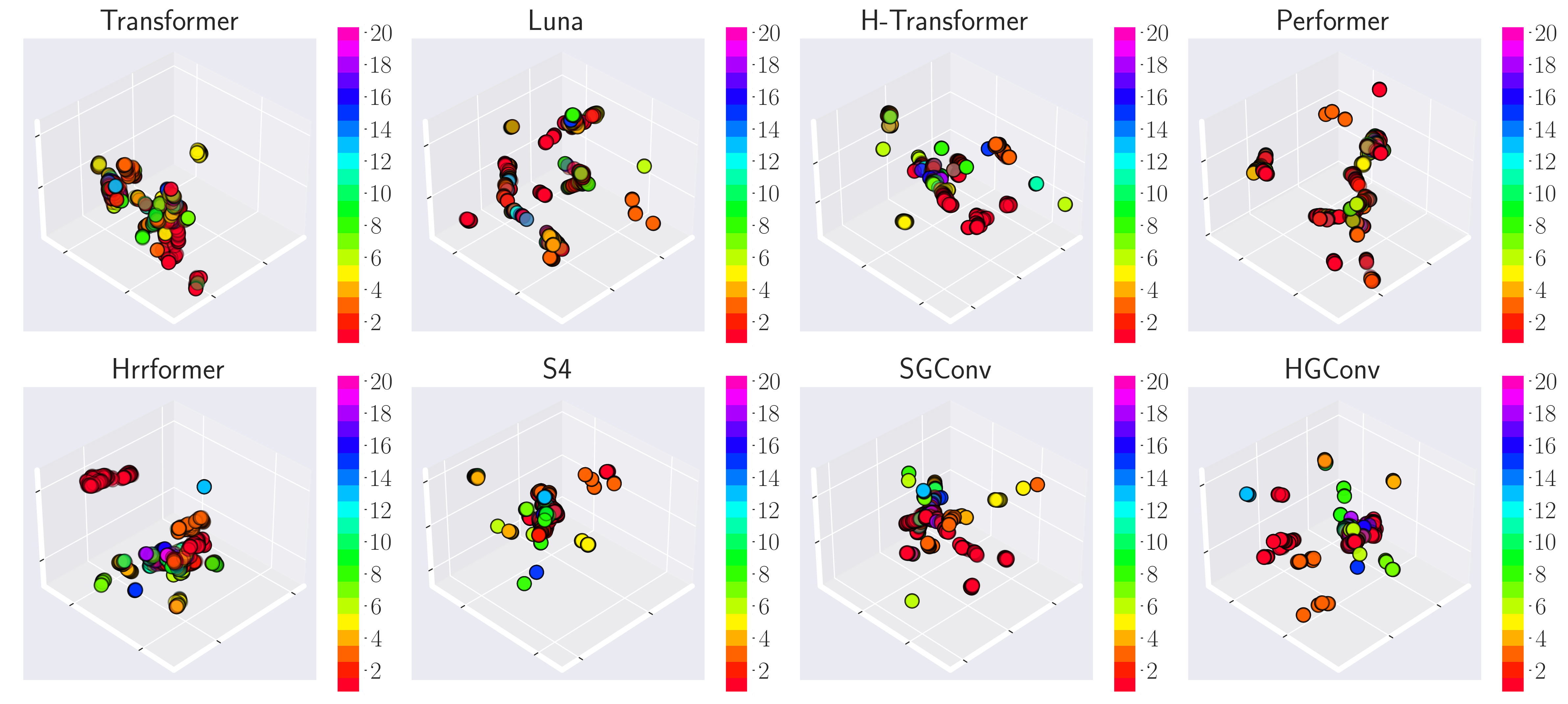}}
\caption{\textsc{Drebin Tar}} 
\label{fig:drebin_tar_umap}
\end{figure*}

The UMAP 3D representations of the output from the penultimate layer of different models for \textsc{Kaggle Raw}, \textsc{Kaggle Asm}, \textsc{Drebin Apk}, and \textsc{Drebin Tar} are presented in \autoref{fig:kaggle_raw_umap}, \autoref{fig:kaggle_asm_umap}, \autoref{fig:drebin_apk_umap}, and \autoref{fig:drebin_tar_umap}, respectively. For \textsc{Kaggle Raw} and \textsc{Asm}, S4, SGConv, Hrrformer, HGConv has quite close accuracy, on the other hand for \textsc{Drebin Apk} and \textsc{Tar}, high variance in accuracy can be noted. Thus, noticeable changes can be observed for \textsc{Drebin Apk} and \textsc{Tar} in \autoref{fig:drebin_apk_umap} and \autoref{fig:drebin_tar_umap} where HGConv has visibly better clusters which makes the final layer classifier predict correctly.

\clearpage
\section{EMBER Comprehensive Results} \label{sec:appendix_d}
\begin{figure*}[!htbp]
\centerline{\includegraphics[width=\columnwidth]{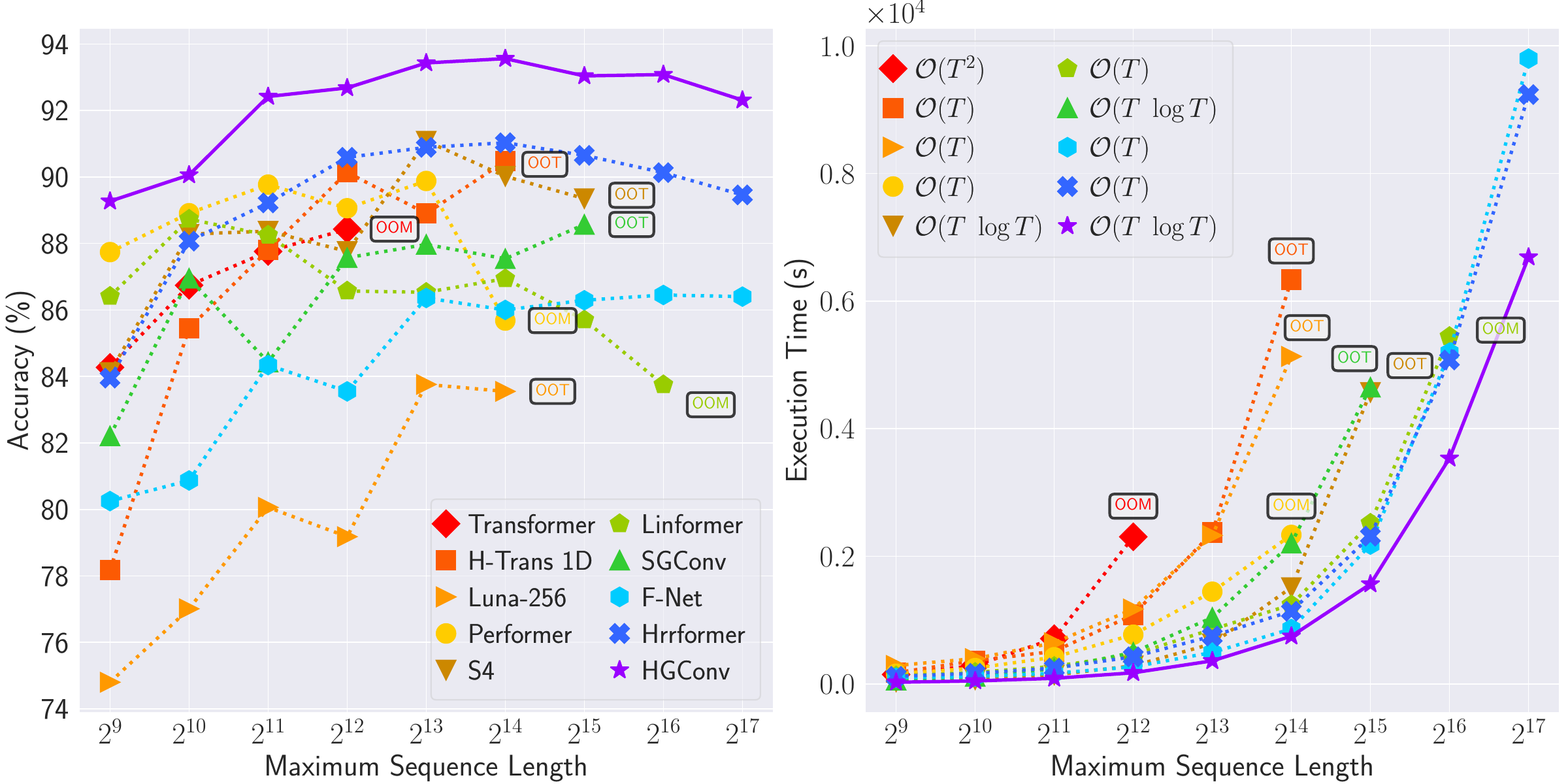}}
\caption{Ember long sequence malware classification. In the figure, OOT and OOM stand for out-of-time (OOT) and memory (OOM) shown for models that face such issues after a particular sequence length.} 
\label{fig:ember}
\end{figure*}

A broader comparison on EMBER classification results is shown in \autoref{fig:ember} with additional models Linformer \citep{linformer}, Performer \citep{performers}, and F-Net \citep{fnet}. The numeric results of each method for different sequence lengths are presented in \autoref{tab:ember_table} where columns are kept empty if it faces OOM or OOT issue. 

\begin{table*}[!htbp]
\centering
\caption{Ember malware classification benchmark. The maximum sequence length ranges from $2^{8}$ to $2^{17}$. The best results are \textbf{boldfaced} and the second-best results are \underline{underlined}.}
\label{tab:ember_table}
\renewcommand{\arraystretch}{1.1}
\resizebox{\textwidth}{!}{
\begin{tabular}{@{}lccccccccccc@{}}
\toprule
\multirow{2}{*}{\textsc{Model}} & \multirow{2}{*}{\textsc{Metrics}} & \multicolumn{10}{c}{\textsc{Maximum Sequence Length}} \\ \cmidrule(l){3-12} 
&  & $\mathrm{256}$ & $\mathrm{512}$ & $\mathrm{1,024}$ & $\mathrm{2,048}$ & $\mathrm{4,096}$ & $\mathrm{8,192}$ & $\mathrm{16,384}$ & $\mathrm{32,768}$ & $\mathrm{65,536}$ & $\mathrm{131,072}$ \\ \toprule
\multirow{2}{*}{\textsc{Transformer} \citep{transformer}} & \textsc{Acc} & 74.87 & 84.27 & 86.74 & 87.76 & 88.43 & OOM & & & & \\
& \textsc{Time} & 101.59 & 146.96 & 286.98 & 708.7 & 2305.28 & OOM & & & & \\ \midrule
\multirow{2}{*}{\textsc{Performer} \citep{performers}} & \textsc{Acc} & 78.0 & \underline{87.74} & 88.91 & \underline{89.77} & 89.06 & 89.88 & 85.68 & OOM & & \\
& \textsc{Time} & 115.77 & 159.59 & 247.02 & 418.1 & 770.75 & 1444.38 & 2334.94 & OOM &  & \\ \midrule
\multirow{2}{*}{\textsc{F-Net} \citep{fnet}} & \textsc{Acc} & 76.42 & 80.25 & 80.87 & 84.34 & 83.55 & 86.36 & 86.00 & 86.29 & 86.45 & 86.40 \\
& \textsc{Time} & 84.84 & 95.58 & 113.2 & 165.77 & \underline{267.21} & \underline{492.44} & \underline{861.48} & \underline{2182.3}0 & 5191.26 & 9800.97 \\ \midrule
\multirow{2}{*}{\textsc{Luna-256} \citep{luna}} & \textsc{Acc} & 70.21 & 74.8 & 77.01 & 80.06 & 79.18 & 83.76 & 83.55 & OOT & & \\
& \textsc{Time} & 243.04 & 287.5 & 395.87 & 643.81 & 1172.35 & 2326.15 & 5132.95 & OOT & & \\ \midrule
\multirow{2}{*}{\textsc{H-Transformer} \citep{h-transformer}} & \textsc{Acc} & 59.59 & 78.17 & 85.45 & 87.8 & 90.14 & 88.9  & 90.48 & OOT & &\\
& \textsc{Time} & 116.6 & 175.04 & 362.41 & 509.63 & 1082.67 & 2371.96  & 6336.37 & OOT & & \\ \midrule
\multirow{2}{*}{\textsc{Hrrformer} \citep{hrrformer}} & \textsc{Acc} & 78.06 & 83.95 & 88.07 & 89.22 & \underline{90.59} & 90.89 & 91.03 & \underline{90.65} & \underline{90.13} & \underline{89.46} \\
& \textsc{Time} & 91.35 & 117.96 & 165.18 & 247.32 & 423.55 & 748.48 & 1138.75 & 2315.62 & \underline{5076.65} & \underline{9237.78} \\ \midrule
\multirow{2}{*}{\textsc{S4} \citep{s4}} & \textsc{Acc} & 75.21 & 84.01 & \underline{89.98} & 89.54 & 88.03 & \underline{91.39} & \underline{91.05} & 90.32 & OOT & \\
& \textsc{Time} & \underline{25.19} & \underline{34.48} & \underline{70.4} & \underline{133.58} & 276.98 & 615.95 & 1484.02 & 4486.83 & OOT & \\ \midrule
\multirow{2}{*}{\textsc{SGConv} \citep{sgconv}} & \textsc{Acc} & 74.45 & 82.21 & 86.95 & 84.43 & 87.57 & 87.97 & 87.54 & 88.57 & OOT & \\
& \textsc{Time} & 28.92 & 54.73 & 113.67 & 235.87 & 493.43 & 1039.98 & 2203.36 & 4648.69 & OOT & \\ \midrule

\multirow{2}{*}{\textbf{\textsc{HGConv}}} & \textsc{Acc} & \textbf{80.66} & \textbf{89.27} & \textbf{91.19} & \textbf{92.42} & \textbf{92.68} & \textbf{93.43} & \textbf{93.56} & \textbf{93.04} & \textbf{93.08} & \textbf{92.31} \\
& \textsc{Time} & \textbf{16.01} & \textbf{24.90} & \textbf{45.16} & \textbf{86.62} & \textbf{173.42} & \textbf{360.81} & \textbf{746.44} & \textbf{1564.69} & \textbf{3536.16} & \textbf{6689.41} \\ \bottomrule
\end{tabular}
}
\end{table*}

\end{document}

%% file: figures/bind_ops.tex
\newcommand{\bind}{%
  \mathrel{%
\begin{tikzpicture}[x=0.75pt,y=0.75pt,yscale=-1,xscale=1]
\draw   (0,5) .. controls (0,2.24) and (2.24,0) .. (5,0) .. controls (7.76,0) and (10,2.24) .. (10,5) .. controls (10,7.76) and (7.76,10) .. (5,10) .. controls (2.24,10) and (0,7.76) .. (0,5) -- cycle ;
\draw    (5,0) -- (1.57,8.57) ;
\draw    (8.47,1.43) -- (5,10) ;
\draw    (0.33,3.33) -- (9.67,3.33) ;
\draw    (0.33,6.67) -- (9.67,6.67) ;
\end{tikzpicture}
  }%
}

%% file: main.bbl
\begin{thebibliography}{}

\bibitem[Ahdritz et~al., 2022]{ahdritz2022openfold}
Ahdritz, G., Bouatta, N., Kadyan, S., Xia, Q., Gerecke, W., O’Donnell, T.~J.,
  Berenberg, D., Fisk, I., Zanichelli, N., Zhang, B., et~al. (2022).
\newblock Openfold: Retraining alphafold2 yields new insights into its learning
  mechanisms and capacity for generalization.
\newblock {\em bioRxiv}, pages 2022--11.

\bibitem[Alam et~al., 2023a]{hrrformer}
Alam, M.~M., Raff, E., Biderman, S., Oates, T., and Holt, J. (2023a).
\newblock Recasting self-attention with holographic reduced representations.
\newblock {\em arXiv preprint arXiv:2305.19534}.

\bibitem[Alam et~al., 2023b]{alam2023generalization}
Alam, M.~M., Raff, E., and Oates, T. (2023b).
\newblock Towards generalization in subitizing with neuro-symbolic loss using
  holographic reduced representations.

\bibitem[Alam et~al., 2022]{pmlr-v162-alam22a}
Alam, M.~M., Raff, E., Oates, T., and Holt, J. (2022).
\newblock Deploying convolutional networks on untrusted platforms using 2{D}
  holographic reduced representations.
\newblock In Chaudhuri, K., Jegelka, S., Song, L., Szepesvari, C., Niu, G., and
  Sabato, S., editors, {\em Proceedings of the 39th International Conference on
  Machine Learning}, volume 162 of {\em Proceedings of Machine Learning
  Research}, pages 367--393. PMLR.

\bibitem[Anderson and Roth, 2018]{ember}
Anderson, H.~S. and Roth, P. (2018).
\newblock Ember: an open dataset for training static pe malware machine
  learning models.
\newblock {\em arXiv preprint arXiv:1804.04637}.

\bibitem[Arp et~al., 2014]{drebin}
Arp, D., Spreitzenbarth, M., Hubner, M., Gascon, H., Rieck, K., and Siemens, C.
  (2014).
\newblock Drebin: Effective and explainable detection of android malware in
  your pocket.
\newblock In {\em Ndss}, volume~14, pages 23--26.

\bibitem[Avsec et~al., 2021]{avsec2021effective}
Avsec, {\v{Z}}., Agarwal, V., Visentin, D., Ledsam, J.~R., Grabska-Barwinska,
  A., Taylor, K.~R., Assael, Y., Jumper, J., Kohli, P., and Kelley, D.~R.
  (2021).
\newblock Effective gene expression prediction from sequence by integrating
  long-range interactions.
\newblock {\em Nature methods}, 18(10):1196--1203.

\bibitem[Ba et~al., 2016]{ln}
Ba, J.~L., Kiros, J.~R., and Hinton, G.~E. (2016).
\newblock Layer normalization.
\newblock {\em arXiv preprint arXiv:1607.06450}.

\bibitem[Botacin et~al., 2021]{Botacin_Galhardo}
Botacin, M., Galhardo~Moia, V.~H., Ceschin, F., Amaral~Henriques, M.~A., and
  Grégio, A. (2021).
\newblock Understanding uses and misuses of similarity hashing functions for
  malware detection and family clustering in actual scenarios.
\newblock {\em Forensic Science International: Digital Investigation},
  38:301220.

\bibitem[Breitinger et~al., 2013]{Breitinger_Astebol_Baier_Busch_2013}
Breitinger, F., Astebol, K.~P., Baier, H., and Busch, C. (2013).
\newblock mvhash-b - a new approach for similarity preserving hashing.
\newblock In {\em Proceedings of the 2013 Seventh International Conference on
  IT Security Incident Management and IT Forensics}, IMF ’13, page 33–44,
  Washington, DC, USA. IEEE Computer Society.

\bibitem[Brown et~al., 2020]{brown2020language}
Brown, T., Mann, B., Ryder, N., Subbiah, M., Kaplan, J.~D., Dhariwal, P.,
  Neelakantan, A., Shyam, P., Sastry, G., Askell, A., et~al. (2020).
\newblock Language models are few-shot learners.
\newblock {\em Advances in neural information processing systems},
  33:1877--1901.

\bibitem[Chen et~al., 2023]{chen2023extending}
Chen, S., Wong, S., Chen, L., and Tian, Y. (2023).
\newblock Extending context window of large language models via positional
  interpolation.
\newblock {\em arXiv preprint arXiv:2306.15595}.

\bibitem[Choromanski et~al., 2020]{performers}
Choromanski, K., Likhosherstov, V., Dohan, D., Song, X., Gane, A., Sarlos, T.,
  Hawkins, P., Davis, J., Mohiuddin, A., Kaiser, L., et~al. (2020).
\newblock Rethinking attention with performers.
\newblock {\em arXiv preprint arXiv:2009.14794}.

\bibitem[Dalla-Torre et~al., 2023]{dalla2023nucleotide}
Dalla-Torre, H., Gonzalez, L., Mendoza-Revilla, J., Carranza, N.~L.,
  Grzywaczewski, A.~H., Oteri, F., Dallago, C., Trop, E., de~Almeida, B.~P.,
  Sirelkhatim, H., et~al. (2023).
\newblock The nucleotide transformer: Building and evaluating robust foundation
  models for human genomics.
\newblock {\em bioRxiv}, pages 2023--01.

\bibitem[Dao, 2023]{dao2023flashattention}
Dao, T. (2023).
\newblock Flashattention-2: Faster attention with better parallelism and work
  partitioning.
\newblock {\em arXiv preprint arXiv:2307.08691}.

\bibitem[Dao et~al., 2022]{dao2022flashattention}
Dao, T., Fu, D., Ermon, S., Rudra, A., and R{\'e}, C. (2022).
\newblock Flashattention: Fast and memory-efficient exact attention with
  io-awareness.
\newblock {\em Advances in Neural Information Processing Systems},
  35:16344--16359.

\bibitem[Dauphin et~al., 2017]{glu}
Dauphin, Y.~N., Fan, A., Auli, M., and Grangier, D. (2017).
\newblock Language modeling with gated convolutional networks.
\newblock In {\em International conference on machine learning}, pages
  933--941. PMLR.

\bibitem[Devlin et~al., 2018]{devlin2018bert}
Devlin, J., Chang, M.-W., Lee, K., and Toutanova, K. (2018).
\newblock Bert: Pre-training of deep bidirectional transformers for language
  understanding.
\newblock {\em arXiv preprint arXiv:1810.04805}.

\bibitem[Du et~al., 2023]{du2023spiking}
Du, Y., Liu, X., and Chua, Y. (2023).
\newblock Spiking structured state space model for monaural speech enhancement.
\newblock {\em arXiv preprint arXiv:2309.03641}.

\bibitem[{E}dward {R}aff et~al., 2019]{pylzjd-proc-scipy-2019}
{E}dward {R}aff, {J}oe {A}urelio, and {C}harles {N}icholas (2019).
\newblock {P}y{L}{Z}{J}{D}: {A}n {E}asy to {U}se {T}ool for {M}achine
  {L}earning.
\newblock In {C}hris {C}alloway, {D}avid {L}ippa, {D}illon {N}iederhut, and
  {D}avid {S}hupe, editors, {\em {P}roceedings of the 18th {P}ython in
  {S}cience {C}onference}, pages 101 -- 106.

\bibitem[Gao et~al., 2021]{eval-harness}
Gao, L., Tow, J., Abbasi, B., Biderman, S., Black, S., DiPofi, A., Foster, C.,
  Golding, L., Hsu, J., Le~Noac'h, A., Li, H., McDonell, K., Muennighoff, N.,
  Ociepa, Chris~Phang, J., Reynolds, L., Schoelkopf, H., Skowron, A., Sutawika,
  L., Tang, E., Thite, A., Wang, B., Wang, K., and Zou, A. (2021).
\newblock A framework for few-shot language model evaluation.

\bibitem[Gu and Dao, 2023]{gu2023mamba}
Gu, A. and Dao, T. (2023).
\newblock Mamba: Linear-time sequence modeling with selective state spaces.
\newblock {\em arXiv preprint arXiv:2312.00752}.

\bibitem[Gu et~al., 2020]{hippo}
Gu, A., Dao, T., Ermon, S., Rudra, A., and R{\'e}, C. (2020).
\newblock Hippo: Recurrent memory with optimal polynomial projections.
\newblock {\em Advances in neural information processing systems},
  33:1474--1487.

\bibitem[Gu et~al., 2021]{s4}
Gu, A., Goel, K., and R{\'e}, C. (2021).
\newblock Efficiently modeling long sequences with structured state spaces.
\newblock {\em arXiv preprint arXiv:2111.00396}.

\bibitem[Hendrycks and Gimpel, 2016]{gelu}
Hendrycks, D. and Gimpel, K. (2016).
\newblock Gaussian error linear units (gelus).
\newblock {\em arXiv preprint arXiv:1606.08415}.

\bibitem[Ioffe and Szegedy, 2015]{bn}
Ioffe, S. and Szegedy, C. (2015).
\newblock Batch normalization: Accelerating deep network training by reducing
  internal covariate shift.
\newblock In {\em International conference on machine learning}, pages
  448--456. pmlr.

\bibitem[Jumper et~al., 2020]{jumper2020alphafold}
Jumper, J., Evans, R., Pritzel, A., Green, T., Figurnov, M., Tunyasuvunakool,
  K., Ronneberger, O., Bates, R., {\v{Z}}{\'\i}dek, A., Bridgland, A., et~al.
  (2020).
\newblock Alphafold 2.
\newblock {\em Fourteenth Critical Assessment of Techniques for Protein
  Structure Prediction; DeepMind: London, UK}.

\bibitem[Katharopoulos et~al., 2020]{linear-trans}
Katharopoulos, A., Vyas, A., Pappas, N., and Fleuret, F. (2020).
\newblock Transformers are rnns: Fast autoregressive transformers with linear
  attention.
\newblock In {\em International Conference on Machine Learning}, pages
  5156--5165. PMLR.

\bibitem[Lee-Thorp et~al., 2021]{fnet}
Lee-Thorp, J., Ainslie, J., Eckstein, I., and Ontanon, S. (2021).
\newblock Fnet: Mixing tokens with fourier transforms.
\newblock {\em arXiv preprint arXiv:2105.03824}.

\bibitem[Li et~al., 2022]{sgconv}
Li, Y., Cai, T., Zhang, Y., Chen, D., and Dey, D. (2022).
\newblock What makes convolutional models great on long sequence modeling?
\newblock {\em arXiv preprint arXiv:2210.09298}.

\bibitem[Lillis et~al., 2017]{Lillis_Breitinger_Scanlon_2017}
Lillis, D., Breitinger, F., and Scanlon, M. (2017).
\newblock Expediting mrsh-v2 approximate matching with hierarchical bloom
  filter trees.
\newblock In {\em 9th EAI International Conference on Digital Forensics and
  Cyber Crime (ICDF2C 2017)}, Prague, Czechia. Springer.

\bibitem[Lin et~al., 2022]{lin2022language}
Lin, Z., Akin, H., Rao, R., Hie, B., Zhu, Z., Lu, W., dos Santos~Costa, A.,
  Fazel-Zarandi, M., Sercu, T., Candido, S., et~al. (2022).
\newblock Language models of protein sequences at the scale of evolution enable
  accurate structure prediction.
\newblock {\em BioRxiv}, 2022:500902.

\bibitem[Linsley et~al., 2018]{pathfinder}
Linsley, D., Kim, J., Veerabadran, V., Windolf, C., and Serre, T. (2018).
\newblock Learning long-range spatial dependencies with horizontal gated
  recurrent units.
\newblock {\em Advances in neural information processing systems}, 31.

\bibitem[Lu et~al., 2023]{lu2023structured}
Lu, C., Schroecker, Y., Gu, A., Parisotto, E., Foerster, J., Singh, S., and
  Behbahani, F. (2023).
\newblock Structured state space models for in-context reinforcement learning.
\newblock {\em arXiv preprint arXiv:2303.03982}.

\bibitem[Ma et~al., 2021]{luna}
Ma, X., Kong, X., Wang, S., Zhou, C., May, J., Ma, H., and Zettlemoyer, L.
  (2021).
\newblock Luna: Linear unified nested attention.
\newblock {\em Advances in Neural Information Processing Systems},
  34:2441--2453.

\bibitem[Maas et~al., 2011]{imdb}
Maas, A., Daly, R.~E., Pham, P.~T., Huang, D., Ng, A.~Y., and Potts, C. (2011).
\newblock Learning word vectors for sentiment analysis.
\newblock In {\em Proceedings of the 49th annual meeting of the association for
  computational linguistics: Human language technologies}, pages 142--150.

\bibitem[McInnes et~al., 2018]{umap}
McInnes, L., Healy, J., and Melville, J. (2018).
\newblock Umap: Uniform manifold approximation and projection for dimension
  reduction.
\newblock {\em arXiv preprint arXiv:1802.03426}.

\bibitem[Menet et~al., 2023]{menet2023mimo}
Menet, N., Hersche, M., Karunaratne, G., Benini, L., Sebastian, A., and Rahimi,
  A. (2023).
\newblock Mimonets: Multiple-input-multiple-output neural networks exploiting
  computation in superposition.
\newblock {\em Advances in Neural Information Processing Systems (NeurIPS)},
  36.

\bibitem[Muennighoff et~al., 2023]{bloomz}
Muennighoff, N., Wang, T., Sutawika, L., Roberts, A., Biderman, S., Le~Scao,
  T., Bari, M.~S., Shen, S., Yong, Z.~X., Schoelkopf, H., Tang, X., Radev, D.,
  Aji, A.~F., Almubarak, K., Albanie, S., Alyafeai, Z., Webson, A., Raff, E.,
  and Raffel, C. (2023).
\newblock Crosslingual generalization through multitask finetuning.
\newblock In {\em Proceedings of the 61st Annual Meeting of the Association for
  Computational Linguistics (Volume 1: Long Papers)}.

\bibitem[Nguyen et~al., 2023]{nguyen2023hyenadna}
Nguyen, E., Poli, M., Faizi, M., Thomas, A., Birch-Sykes, C., Wornow, M.,
  Patel, A., Rabideau, C., Massaroli, S., Bengio, Y., et~al. (2023).
\newblock Hyenadna: Long-range genomic sequence modeling at single nucleotide
  resolution.
\newblock {\em arXiv preprint arXiv:2306.15794}.

\bibitem[Nolet et~al., 2021]{Nolet2021}
Nolet, C.~J., Lafargue, V., Raff, E., Nanditale, T., Oates, T., Zedlewski, J.,
  and Patterson, J. (2021).
\newblock Bringing umap closer to the speed of light with gpu acceleration.
\newblock {\em Proceedings of the AAAI Conference on Artificial Intelligence},
  35(1):418–426.

\bibitem[Oliver et~al., 2013]{Oliver_Cheng_Chen_2013}
Oliver, J., Cheng, C., and Chen, Y. (2013).
\newblock Tlsh -- a locality sensitive hash.
\newblock In {\em 2013 Fourth Cybercrime and Trustworthy Computing Workshop},
  page 7–13. IEEE.

\bibitem[Panconesi et~al., 2015]{kaggle}
Panconesi, A., Marian, Cukierski, W., and Committee, W. B.~C. (2015).
\newblock Microsoft malware classification challenge (big 2015).

\bibitem[Peng et~al., 2023a]{peng2023rwkv}
Peng, B., Alcaide, E., Anthony, Q., Albalak, A., Arcadinho, S., Cao, H., Cheng,
  X., Chung, M., Grella, M., GV, K.~K., et~al. (2023a).
\newblock Rwkv: Reinventing rnns for the transformer era.
\newblock {\em arXiv preprint arXiv:2305.13048}.

\bibitem[Peng et~al., 2023b]{peng2023yarn}
Peng, B., Quesnelle, J., Fan, H., and Shippole, E. (2023b).
\newblock Yarn: Efficient context window extension of large language models.
\newblock {\em arXiv preprint arXiv:2309.00071}.

\bibitem[Plate, 1995]{hrr}
Plate, T.~A. (1995).
\newblock Holographic reduced representations.
\newblock {\em IEEE Transactions on Neural networks}, 6(3):623--641.

\bibitem[Poli et~al., 2023]{poli2023hyena}
Poli, M., Massaroli, S., Nguyen, E., Fu, D.~Y., Dao, T., Baccus, S., Bengio,
  Y., Ermon, S., and R{\'e}, C. (2023).
\newblock Hyena hierarchy: Towards larger convolutional language models.
\newblock {\em arXiv preprint arXiv:2302.10866}.

\bibitem[Radev et~al., 2013]{aan}
Radev, D.~R., Muthukrishnan, P., Qazvinian, V., and Abu-Jbara, A. (2013).
\newblock The acl anthology network corpus.
\newblock {\em Language Resources and Evaluation}, 47(4):919--944.

\bibitem[Raff et~al., 2021]{Raff2021}
Raff, E., Fleshman, W., Zak, R., Anderson, H.~S., Filar, B., and McLean, M.
  (2021).
\newblock Classifying sequences of extreme length with constant memory applied
  to malware detection.
\newblock {\em Proceedings of the {AAAI} Conference on Artificial
  Intelligence}, 35(11):9386--9394.

\bibitem[Raff and Nicholas, 2017a]{10.1145/3097983.3098111}
Raff, E. and Nicholas, C. (2017a).
\newblock An alternative to ncd for large sequences, lempel-ziv jaccard
  distance.
\newblock In {\em Proceedings of the 23rd ACM SIGKDD International Conference
  on Knowledge Discovery and Data Mining}, KDD '17, page 1007–1015, New York,
  NY, USA. Association for Computing Machinery.

\bibitem[Raff and Nicholas, 2017b]{raff2017malware}
Raff, E. and Nicholas, C. (2017b).
\newblock Malware classification and class imbalance via stochastic hashed
  lzjd.
\newblock In {\em Proceedings of the 10th ACM Workshop on Artificial
  Intelligence and Security}, pages 111--120.

\bibitem[Raff and Nicholas, 2018a]{RAFF201834}
Raff, E. and Nicholas, C. (2018a).
\newblock Lempel-ziv jaccard distance, an effective alternative to ssdeep and
  sdhash.
\newblock {\em Digital Investigation}, 24:34--49.

\bibitem[Raff and Nicholas, 2020]{Raff_Nicholas_2020}
Raff, E. and Nicholas, C. (2020).
\newblock A survey of machine learning methods and challenges for windows
  malware classification.
\newblock In {\em NeurIPS 2020 Workshop: ML Retrospectives, Surveys \&
  Meta-Analyses (ML-RSA)}.

\bibitem[Raff et~al., 2020]{BWMD}
Raff, E., Nicholas, C., and McLean, M. (2020).
\newblock {A New Burrows Wheeler Transform Markov Distance}.
\newblock In {\em The Thirty-Fourth AAAI Conference on Artificial
  Intelligence}.

\bibitem[Raff and Nicholas, 2018b]{Raff_Nicholas_2018}
Raff, E. and Nicholas, C.~K. (2018b).
\newblock Lempel-ziv jaccard distance, an effective alternative to ssdeep and
  sdhash.
\newblock {\em Digital Investigation}.

\bibitem[Raffel et~al., 2020]{raffel2020exploring}
Raffel, C., Shazeer, N., Roberts, A., Lee, K., Narang, S., Matena, M., Zhou,
  Y., Li, W., and Liu, P.~J. (2020).
\newblock Exploring the limits of transfer learning with a unified text-to-text
  transformer.
\newblock {\em The Journal of Machine Learning Research}, 21(1):5485--5551.

\bibitem[Raji et~al., 2021]{raji2021ai}
Raji, I.~D., Denton, E., Bender, E.~M., Hanna, A., and Paullada, A. (2021).
\newblock Ai and the everything in the whole wide world benchmark.
\newblock In {\em Thirty-fifth Conference on Neural Information Processing
  Systems Datasets and Benchmarks Track (Round 2)}.

\bibitem[Romero and Zeghidour, 2023]{romero2023dnarch}
Romero, D.~W. and Zeghidour, N. (2023).
\newblock Dnarch: Learning convolutional neural architectures by
  backpropagation.
\newblock {\em arXiv preprint arXiv:2302.05400}.

\bibitem[Roussev, 2009]{Roussev_2009}
Roussev, V. (2009).
\newblock Building a better similarity trap with statistically improbable
  features.
\newblock In {\em Proceedings of the 42Nd Hawaii International Conference on
  System Sciences}, HICSS ’09, page 1–10, Washington, DC, USA. IEEE
  Computer Society.

\bibitem[Rozi{\`e}re et~al., 2023]{roziere2023code}
Rozi{\`e}re, B., Gehring, J., Gloeckle, F., Sootla, S., Gat, I., Tan, X.~E.,
  Adi, Y., Liu, J., Remez, T., Rapin, J., et~al. (2023).
\newblock Code llama: Open foundation models for code.
\newblock {\em arXiv preprint arXiv:2308.12950}.

\bibitem[Saul et~al., 2023]{pmlr-v187-saul23a}
Saul, R., Alam, M.~M., Hurwitz, J., Raff, E., Oates, T., and Holt, J. (2023).
\newblock Lempel-ziv networks.
\newblock In Antorán, J., Blaas, A., Feng, F., Ghalebikesabi, S., Mason, I.,
  Pradier, M.~F., Rohde, D., Ruiz, F. J.~R., and Schein, A., editors, {\em
  Proceedings on "I Can't Believe It's Not Better! - Understanding Deep
  Learning Through Empirical Falsification" at NeurIPS 2022 Workshops}, volume
  187 of {\em Proceedings of Machine Learning Research}, pages 1--11. PMLR.

\bibitem[Tay et~al., 2020]{lra}
Tay, Y., Dehghani, M., Abnar, S., Shen, Y., Bahri, D., Pham, P., Rao, J., Yang,
  L., Ruder, S., and Metzler, D. (2020).
\newblock Long range arena: A benchmark for efficient transformers.
\newblock {\em arXiv preprint arXiv:2011.04006}.

\bibitem[Team, 2023]{MosaicML2023Introducing}
Team, M.~N. (2023).
\newblock Introducing mpt-30b: Raising the bar for open-source foundation
  models.
\newblock Accessed: 2023-06-22.

\bibitem[Vaswani et~al., 2017]{transformer}
Vaswani, A., Shazeer, N., Parmar, N., Uszkoreit, J., Jones, L., Gomez, A.~N.,
  Kaiser, {\L}., and Polosukhin, I. (2017).
\newblock Attention is all you need.
\newblock {\em Advances in neural information processing systems}, 30.

\bibitem[Wang et~al., 2020]{linformer}
Wang, S., Li, B.~Z., Khabsa, M., Fang, H., and Ma, H. (2020).
\newblock Linformer: Self-attention with linear complexity.
\newblock {\em arXiv preprint arXiv:2006.04768}.

\bibitem[Winter et~al., 2013]{Winter_Schneider_Yannikos_2013}
Winter, C., Schneider, M., and Yannikos, Y. (2013).
\newblock F2s2: Fast forensic similarity search through indexing piecewise hash
  signatures.
\newblock {\em Digital Investigation}, 10(4):361–371.

\bibitem[Yao et~al., 2023]{yao2023deepspeed}
Yao, Z., Wu, X., Li, C., Zhang, M., Qi, H., Ruwase, O., Awan, A.~A.,
  Rajbhandari, S., and He, Y. (2023).
\newblock Deepspeed-visualchat: Multi-round multi-image interleave chat via
  multi-modal causal attention.
\newblock {\em arXiv preprint arXiv:2309.14327}.

\bibitem[Zaheer et~al., 2020]{bigbird}
Zaheer, M., Guruganesh, G., Dubey, K.~A., Ainslie, J., Alberti, C., Ontanon,
  S., Pham, P., Ravula, A., Wang, Q., Yang, L., et~al. (2020).
\newblock Big bird: Transformers for longer sequences.
\newblock {\em Advances in Neural Information Processing Systems},
  33:17283--17297.

\bibitem[Zama and Schwenker, 2023]{zama2023ecg}
Zama, M.~H. and Schwenker, F. (2023).
\newblock Ecg synthesis via diffusion-based state space augmented transformer.
\newblock {\em Sensors}, 23(19):8328.

\bibitem[Zhu and Soricut, 2021]{h-transformer}
Zhu, Z. and Soricut, R. (2021).
\newblock H-transformer-1d: Fast one-dimensional hierarchical attention for
  sequences.
\newblock {\em arXiv preprint arXiv:2107.11906}.

\end{thebibliography}
